# Spatiotemporal modeling of mid-infrared photoluminescence from terbium (iii) ion doped chalcogenide-selenide multimode fibers

**Slawomir Sujecki[1,2]\*, Lukasz Sojka[1], Zhuoqi Tang[2], Dinuka Jayasuriya[2], David Furniss[2], Emma Barney[2], Trevor Benson[2] and Angela Seddon[2,]**

[1] Telecommunications and Teleinformatics Department, Wroclaw University of Science and Technology, Wroclaw, Poland; slawomir.sujecki@pwr.edu.pl

[2] Mid-Infrared Photonics Group, George Green Institute for Electromagnetics Research, Faculty of Engineering, The University of Nottingham, Nottingham, UK; angela.seddon@nottingham.ac.uk

\* Correspondence: slawomir.sujecki@pwr.edu.pl; Tel.: +48-71-3204588

**Abstract:** In this contribution a numerical model is developed to study the time dynamics of photoluminescence emitted by $Tb^{3+}$ doped multimode chalcogenide-selenide glass fibers pumped by laser light at approximately 2 μm. The model consists of a set of partial differential equations (PDEs), which describe the temporal and spatial evolution of the photon density and level populations within the fiber. In order to solve numerically the PDEs a Method of Lines is applied. The modeling parameters are extracted from measurements and from data available in the literature. The numerical results obtained support experimental observations. In particular, the developed model reproduces the discrepancies that are observed between the photoluminescence decay curves obtained from different points along the fiber. The numerical analysis is also used to explain the source of these discrepancies.

## 1. Introduction

Mid-infrared (MIR) light has numerous applications in medicine, biology, agriculture, defense and security.

**Foundation item:** This Project has received funding from the European Union's Horizon 2020 research and innovation programme under the Marie Skłodowska-Curie grant agreement No. 665778 (National Science Centre, Poland, Polonez Fellowship 2016/21/P/ST7/03666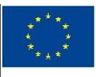. The authors wish to thank also Wroclaw University of Science and Technology (statutory activity) for financial support.

\*corresponding author email address: slawomir.sujecki@pwr.edu.pl, phone: 0048 71 320 4588

Therefore in many laboratories intensive research is being carried out to develop MIR light technology in order to achieve a similar level of technological maturity as in the case of near-infrared and visible light wavelength ranges. One of the key elements of almost any device operating within the MIR part of the spectrum is the light source. There are several types of light source that have been developed for the generation of MIR light. These include the Globar©, quantum cascade lasers, inter-band cascade lasers, gas lasers, Raman lasers, optical parametric amplifiers, supercontinuum sources, lanthanide ion doped fiber lasers and fiber lanthanide ion doped spontaneous emission sources. This contribution is focused on the numerical analysis of the latter lanthanide ion doped spontaneous emission sources (LIDSES), which use chalcogenide glass technology and lanthanide ion doping. Chalcogenide glass LIDSES have been so far successfully implemented in optical sensors and sensor systems operating within the MIR wavelength range [1, 2]. The operating wavelengths of LIDSES using chalcogenide glass can reach at least 5500 nm [3]. This technological evolution towards longer emitted wavelengths was made possible by very rapid progress in chalcogenide glass technology, which has taken place over the past decade [4-8]. This technological progress has been accompanied by the development of simulation and design tools [9-13]. Lanthanide ion doped chalcogenide glass fibers have been successfully fabricated [14-18]. The luminescence from fabricated lanthanide ion doped chalcogenide glasses has been measured independently in several laboratories [19-23].

In this paper a spatiotemporal model is developed for studying the photoluminescence emitted from lanthanide ion doped chalcogenide-selenide glass multimode fibers. The modeling parameters are derived experimentally from lanthanide ion doped chalcogenide-selenide bulk glass and fiber samples. The set of partial differential equations which describes the spatiotemporal dynamics of photon populations and ionic level, electronic state populations is solved numerically using the method of lines. In particular, spatiotemporal dynamics within a terbium (III) ion doped chalcogenide-selenide glass multimode fiber are studied. The results obtained are used to gain a better understanding of the differences between photoluminescence decay curves emitted from various points along the fiber. This study also complements similar studies performed for bulk samples [24-27]. It is noted however that the numerical method developed in this contribution is specifically tailored for studying the photoluminesce emitted by fibers.

**2. Methods and Materials**

For the experimental work, 1000 ppmw (parts per million by weight) $Tb^{3+}$-doped GeAsGaSe bulk glass has been made by the melt-quenching route [28].

Fig.1 and Fig.2 show schematic diagrams of the measurement setups used for recording of photoluminescence spectra and photoluminescence lifetime, respectively. For spectrum measurements (Fig.1) the signal was collected from the fiber side and delivered to monochromator entrance using a set of calcium fluoride lenses. The incident pump light was focused on one end of the fiber using a set of lenses. A Mercury Cadmium Telluride (MCT) photodetector PVI-4TE6 (Vigo Systems) recorded the MIR light intensity. The signal from the MCT photodetector was delivered to a lock-in amplifier together with the reference signal from the chopper driver. Post processing of the results was carried out by a personal computer connected with a data acquisition card. In the case of the photoluminescence decay curves, the pump laser light was modulated by the chopper before entering one end of the fiber (Fig.2). The signal was collected from the side of the fiber and imaged onto the MCT photodetector through a monochromator. The signal from the MCT photodetector was amplified, electrically filtered and supplied to an oscilloscope. In order to suppress the noise, an averaging over several thousand process realizations was performed. In both setups filters were used to suppress the residual pump signal.

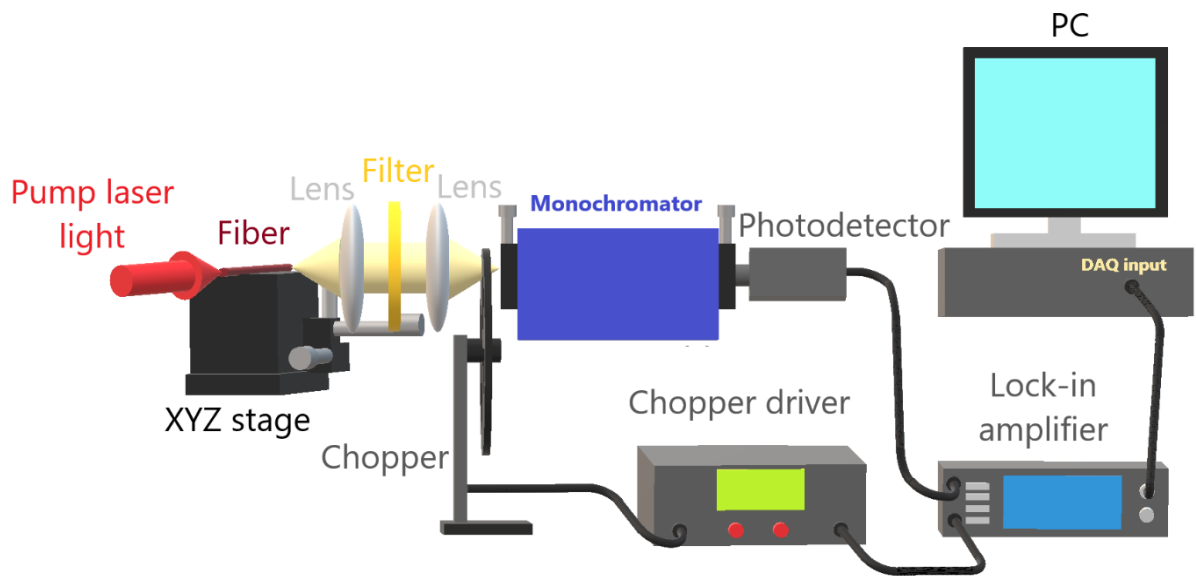

**Figure 1.** Schematic diagram of the photoluminescence spectrum measurement with the chalcogenide fiber pumped at one end using a laser diode and the photoluminescence collected from the side of the fiber.

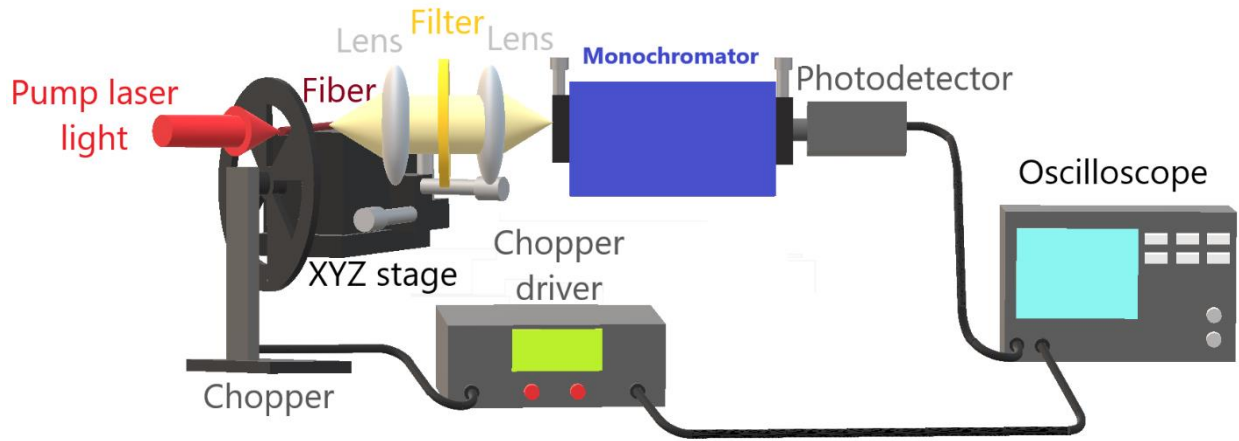

**Figure 2.** Schematic diagram illustrating the photoluminescence decay measurement with the chalcogenide fiber pumped at one end using a laser diode and the photoluminescence collected from the side of the fiber.

Next the outline of the derivation of the main equations describing the dynamics of the photoluminescence studied is given. These equations are solved numerically in the next section. A selenide chalcogenide-selenide glass fiber doped with trivalent terbium ions is considered. Figure 3 shows the energy level diagram for terbium ions doped into a selenide chalcogenide glass. In mid infrared (MIR) applications level 3 is used for pumping. It is assumed that all electronic levels forming level 3 are thermally coupled. MIR light is generated through the interaction of the photon field with the electronic transition 2-1. It was shown experimentally by a number of independent research teams that for a selenide-chalcogenide glass host the transition 2-1 is radiative, unlike the transition 3-2, which is marked with a grey arrow in Fig.3, which is predominantly non-radiative. For terbium the available pumping wavelengths include 1900 nm, 2000 nm, 2300 nm, 3000 nm and 4500 nm. In this study a home-made thulium fiber laser was developed which operated at approximately 2010 nm and was used to pump into level $^7F_2$ (Fig.3).

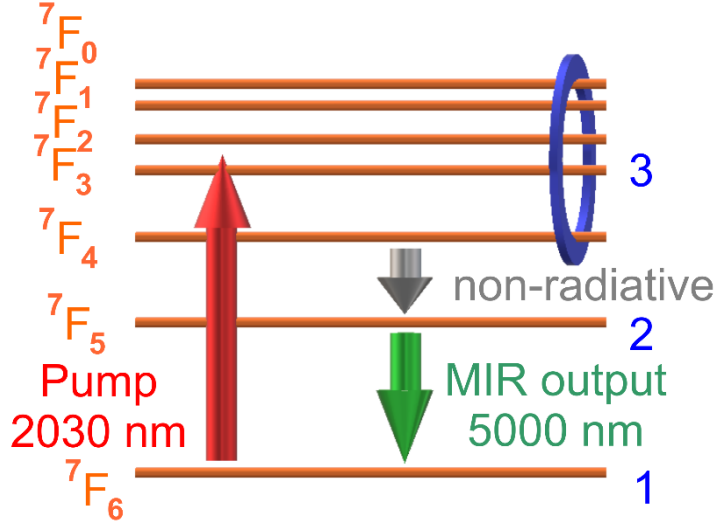

**Figure 3.** Energy level diagram for a terbium ion doped into selenide chalcogenide glass

The rate equations approach applied to the energy levels system presented in Fig.3 yields a set of three coupled differential equations:

$$\frac{\partial N_3}{\partial t} = \left(\sigma_{31a} N_1 - \sigma_{31e} N_3\right)\phi_p - \frac{N_3}{\tau_{32nr}} - \frac{N_3}{\tau_{3r}} \tag{1a}$$

$$\frac{\partial N_2}{\partial t} = \left(\overline{\sigma}_{21a} N_1 - \overline{\sigma}_{21e} N_2\right)\phi_s + \frac{N_3}{\tau_{32nr}} - \frac{N_2}{\tau_{2r}} - \frac{N_2}{\tau_{21nr}} + \frac{\beta_{32} N_3}{\tau_{3r}} \tag{1b}$$

$$\frac{\partial N_1}{\partial t} = -\left(\sigma_{31a} N_1 - \sigma_{31e} N_3\right)\phi_p - \left(\overline{\sigma}_{21a} N_1 - \overline{\sigma}_{21e} N_3\right)\phi_s + \frac{N_2}{\tau_{2r}} + \frac{N_2}{\tau_{21nr}} \frac{\beta_{31} N_3}{\tau_{3r}} \tag{1c}$$

In equation (1) $\sigma_{31a}$ and $\sigma_{31e}$ are the relevant values of the absorption and emission cross sections for the pump wave, respectively, while $\sigma_{21a}$ and $\sigma_{21e}$ give the values of absorption and emission cross sections for the signal wave. $\tau_{2r}$ and $\tau_{3r}$ are radiative lifetimes of level 2 and 3 respectively while $\tau_{32nr}$ and $\tau_{21nr}$ give the relevant life times for phonon assisted transitions. The coefficients $\beta_{32}$ and $\beta_{31}$ give the branching ratios for transitions 3-2 and 3-1, respectively. Additionally, a consistency condition:

$$N_1 + N_2 + N_3 = N_{Tb} \tag{2}$$

imposes that the sum of level population is equal to the total terbium ion concentration $N_{Tb}$.

If the fiber is aligned with the z axis (Fig.4) the power distributions for the pump and signal waves satisfy the partial differential equations:

$$\left(\frac{\partial}{\partial z}+\frac{1}{v_g}\frac{\partial}{\partial t}\right)P_p^+ = -\left(\sigma_{31a}N_1-\sigma_{31e}N_3\right)P_p^+ - \alpha_p P_p^+$$

$$\left(-\frac{\partial}{\partial z}+\frac{1}{v_g}\frac{\partial}{\partial t}\right)P_p^- = -\left(\sigma_{31a}N_1-\sigma_{31e}N_3\right)P_p^- - \alpha_p P_p^-$$

(3a)

$$\left(\frac{\partial}{\partial z}+\frac{1}{v_g}\frac{\partial}{\partial t}\right)P_s^+ = -\left(\overline{\sigma}_{21a}N_1-\overline{\sigma}_{21e}N_2\right)P_s^+ - \overline{\alpha}_s P_s^+ + \gamma_{sp}N_2$$

$$\left(-\frac{\partial}{\partial z}+\frac{1}{v_g}\frac{\partial}{\partial t}\right)P_s^- = -\left(\overline{\sigma}_{21a}N_1-\overline{\sigma}_{21e}N_2\right)P_s^- - \overline{\alpha}_s P_s^- + \gamma_{sp}N_2$$

(3b)

where '+' and '-' refer to forward and backward travelling waves, respectively while, $\alpha_s$ gives the attenuation coefficient for the signal wave, which is averaged with respect to the signal spectrum [1]. The averaged value is marked with an overstrike symbol and was also applied consistently to the signal emission and absorption cross-sections for the signal wave. $\gamma_{sp} = \eta v_g / A*h*f_s$ is the spontaneous emission coupling factor, which was calculated following the method outlined in [1] while $v_g$ is the group velocity of light. The pump and signal fluxes $\phi_p$ and $\phi_s$ are related to the respective power values via: $P_p = \phi_p*h*f_p*A$ and $P_s = \phi_s*h*f_s*A$ where A is the fiber cross sectional area, h is Planck's constant and $f_p$ and $f_s$ are the pump and signal frequencies, respectively. The equations 3 are complemented by the boundary conditions applied at both ends of the fiber:

$$P_p^+(z=0) = R_p(z=0)P_p^-(z=0) + (1-R_p(z=0))P_{pump}$$
$$P_p^-(z=L) = R_p(z=L)P_p^+(z=L)$$
$$P_s^+(z=0) = R_s(z=0)P_s^-(z=0)$$
$$P_s^-(z=L) = R_s(z=L)P_s^+(z=L)$$

(4)

where $R_p$ and $R_s$ are the values for reflectivity for pump and signal waves, respectively. $P_{pump}$ gives the value of the incident pump power and $L$ is the fiber length. The equations 1 and 2 subject to boundary conditions given by 4 are solved using the Method of Lines (MoL). In the MoL the partial differential equations 1 and 2 are converted into a set of ordinary differential equations at a set of sampling points (Fig.4) by applying the finite difference approximations to the partial derivatives with respect to z. The details of the MoL are given in [29].

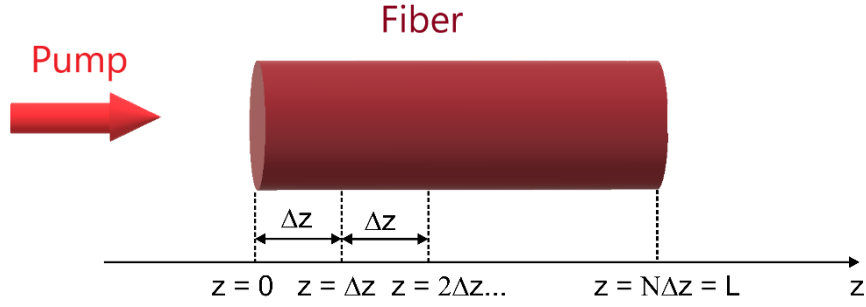

**Figure 4.** Photon flux amplification via a stimulated emission process

In the next section the MoL is applied to calculate the dependence of the photon distributions and level populations on time and z spatial variable. The numerically obtained results are compared with the experimentally measured ones.

## 3. Results and Discussion

The modeling parameters were extracted from experimental measurements described in [21] whilst the multiphonon transition lifetimes were extracted from the experimental data by making a least squares fit under the assumption that the maximum intrinsic phonon energy of $Tb^{3+}$-doped Ge-As-Ga-Se is ~ 300 cm$^{-1}$. The $Tb^{3+}$ ion concentration is $1.65 \times 10^{25}$/m$^3$. The assumed averaged attenuation coefficient for the signal wave is 10 m$^{-1}$ (due to SeH impurities [30]). The weighted signal emission and absorption cross sections are $6.3 \times 10^{-25}$ m$^2$ and $6.7 \times 10^{-25}$ m$^2$, respectively. The pump emission and absorption cross sections are $0 \times 10^{-25}$ m$^2$ and $6.46 \times 10^{-25}$ m$^2$, respectively. The assumed zero value of the pump emission cross section reflects the fact that due to Boltzmann distribution the population of level $^7F_3$ will be much smaller than the population of level $^7F_4$ whilst level $^7F_4$ will in turn be strongly depleted due to the predominantly non-radiative transition between levels 3 and 2, which is characterized by a short lifetime [21]. The reflectivity at the fiber end for chalcogenide selenide glass is assumed to be 0.2 [31] while the spontaneous emission coupling coefficient $\eta = 0.3$ [32]. The lifetimes of level 3 and 2 (Fig.3) are 5.8 ms and 13.1 ms, respectively while the branching ratio of level 3-2 transition is 0.11. The multiphoton transition lifetime for 3-2 transition is assumed to be 0.012 ms. The fiber was unstructured with 320 μm outside diameter. Fig.5 compares the photoluminescence spectrum measured around 3000 nm (due to $^7F_4 - ^7F_6$ transition) and around 4800 nm (due to $^7F_5 - ^7F_6$ transition). These results confirm that for the host glass considered the level 3 is depleted mainly through non-radiative transitions. This is because the observed 3000 nm photoluminescence is very weak compared with the 4800 nm luminescence and is hardly distinguishable from the background noise. It is noted that in other chalcogenide glass hosts this does not have to be the case since an experimental verification of this fact was necessary [19].

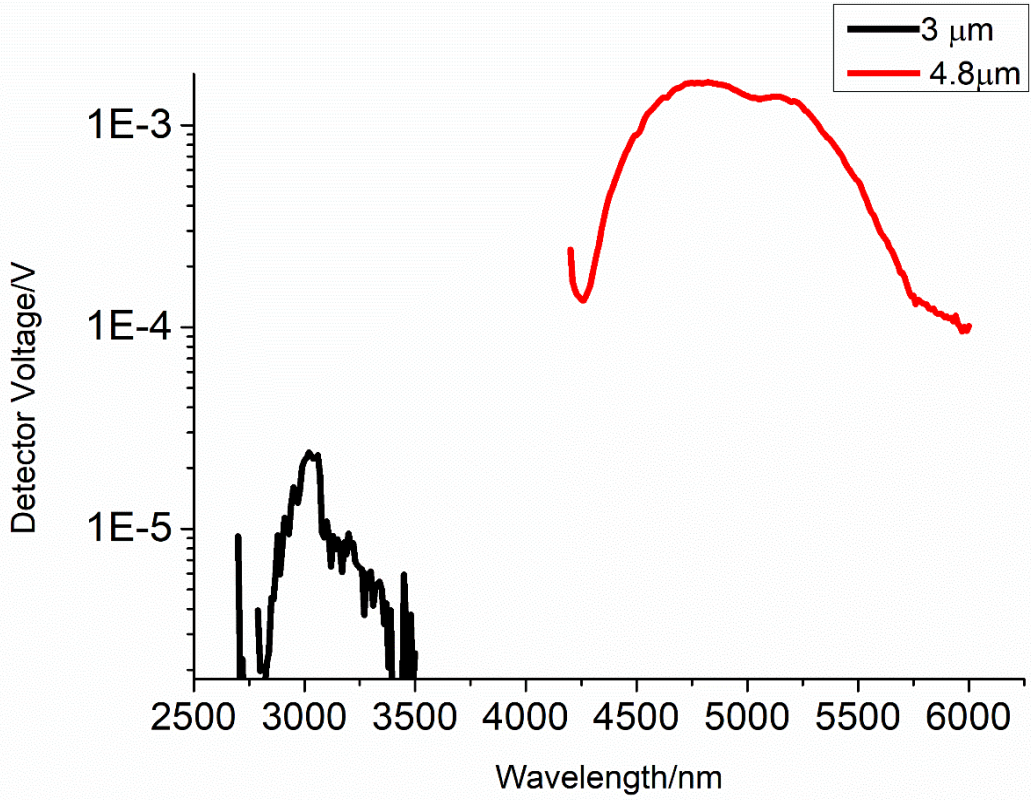

**Figure 5.** Measured photoluminescence collected at 3 μm and 4.8 μm using the experimental setup shown in Fig.1.

Fig.6 shows a comparison of numerically calculated and experimentally measured photoluminescence decay around 4800 nm collected from the side of the fiber at either the pumped fiber end (z = 0, Fig.4) or the other end (z = L = 5 cm), cf. the inset of Fig.6. The numerically calculated curves were obtained from the normalized decay curves of the level 2 population $N_2$. The results from Fig.6 show that the numerical model is able to reproduce the experimentally observed discrepancies between the photoluminescence decay curves collected from various points along the fiber. In order to obtain the good agreement between the experiment and modeling the pump loss coefficient $\alpha_p$ had to be set to 40 m$^{-1}$ whilst the multiphonon transition lifetime for the 2-1 transition, $\tau_{21nr}$, was assumed equal to 15 ms. It is noted that it is not possible to separate the contribution of the multiphonon transition from the radiative transition by observing and analyzing the $N_2$ decay. One can only know the combined effect i.e. extract the time constant $1/(1/\tau_{2r}+1/\tau_{21nr})$. Here, in agreement with the results of Judd-Ofelt analysis published in [21], it was assumed that $\tau_{2r}$ = 13.1 ms whilst $\tau_{21nr}$ was used to tune the numerically calculated decay rate of $N_2$ so that the agreement with the experimental results is achieved. (Please note that all stated fiber parameters were kept constant in all simulation results presented). During the initial few milliseconds the $N_2$ dynamics is strongly dependent on the rate at which the pump power is reduced by being gradually blocked by the chopper blade. This effect of the pump beam modulation by the chopper blade was modelled by assuming a cosine squared time dependence: $\cos^2(\pi*t/2*\tau)$ with the time constant $\tau$ = 2.5 ms. The maximum coupled pump power was assumed to be equal to 100 mW, which is most likely underestimating the true value, considering that the measured incident pump power was approximately 180 mW whilst approximately 20 % loss should be attributed to Fresnel loss and some other fraction of the pump power was lost due to an imperfect beam alignment with the fiber end. It is noted that subsequent simulations, not included in this contribution, showed the calculated $N_2$ time decay curve shapes were not visibly changing upon increasing the fiber coupled pump power up to 180 mW.

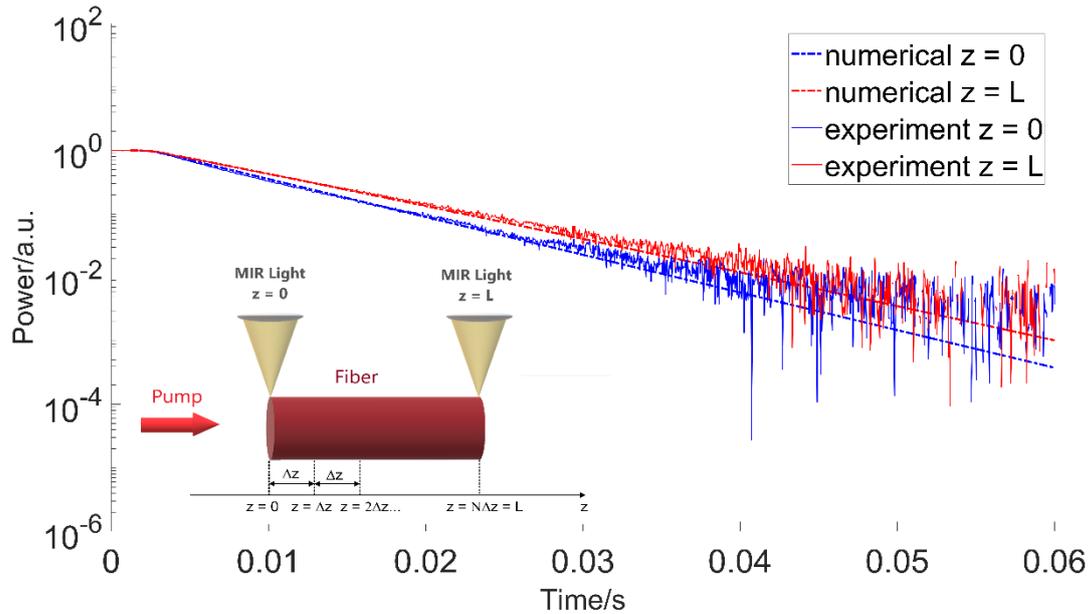

**Figure 6.** Comparison of numerically calculated and experimentally measured photoluminescence decay at 4600 nm collected from the side of the fiber at either the pumped fiber end (z = 0, Fig.4) or the other end (z = L). The assumed fiber coupled pump power is 100 mW.

Considering that the results of Fig.6 show a good agreement of the model results with the experiment it is important to understand its origin. Especially, it is interesting to explore why the model predicts different decay curves when the light is collected from various positions along the fiber.

The dynamics of the level 2 population $N_2$ is governed primarily by equation (1b). One can conclude by inspection that the equation (1b) would predict the correct decay curves if the terms $N_2/\tau_{2r}$ and $N_2/\tau_{21nr}$ appearing on the right hand side of equation (1b) were much larger than other terms. Therefore in Table 1 the various terms appearing on the right hand side of equation (1b) are identified and labeled. In Fig.7 the term values at selected z positions (z = 0 and z = L) calculated numerically are compared as functions of time.

Table 1. Labels of terms appearing on the right hand side of equation (1b)

| Term label | Term |
|---|---|
| $t_1$ | $-\bar{\sigma}_{21e} N_2 \phi_s$ |
| $t_2$ | $\bar{\sigma}_{21a} N_1 \phi_s$ |
| $t_3$ | $\dfrac{\beta_{32} N_3}{\tau_{3r}}$ |
| $t_4$ | $\dfrac{N_3}{\tau_{32nr}}$ |
| $t_5$ | $-\dfrac{N_2}{\tau_{2r}} - \dfrac{N_2}{\tau_{21nr}}$ |

The results presented in Fig.7 show that the terms $t_1$ and $t_3$ are negligibly small when compared with other terms. Term $t_4$ is only relevant during the first few milliseconds, i.e. during the switching off the pump power beam by the chopper. For other time values term $t_4$ is negligibly small. The only terms left are $t_5$ and $t_2$. Term $t_2$ is smaller than $t_5$ but is clearly making a noticeable contribution and thus modifies the dynamics of $N_2$. Also it can be observed that at $z = 0$ term $t_2$ makes a relatively smaller contribution than at $z = L$, which explains the discrepancies in the $N_2$ time dependence recorded at $z = 0$ and $z = L$. Since $t_2$ contribution at $z = 0$ is relatively smaller than at $z = L$ one would expect that the time constant extracted at $z = 0$ is nearer to the expected value of $1/(1/\tau_{2r}+1/\tau_{21nr})$ than the one extracted from the results recorded at $z = L$. In order to verify this conclusion Fig.8 shows the time dependence of the implicit time constant obtained numerically by calculating the derivative of logarithm of $N_2$ from the numerically obtained results. The value of the derivative was approximated using first order backward finite difference. If the term $t_5$ was dominant the calculation of the derivative from logarithm of $N_2$ would yield the expected value of $1/(1/\tau_{2r}+1/\tau_{21nr})$, which is given in Fig.8 as the 'reference' value. This one would expect at least to be the case for sufficiently large values of time when all other effects, represented by terms $t_1$ – $t_4$, e.g. stimulated emission and absorption, become negligibly small when compared with the effect of spontaneous emission and multiphonon assisted transitions, i.e. term $t_5$.

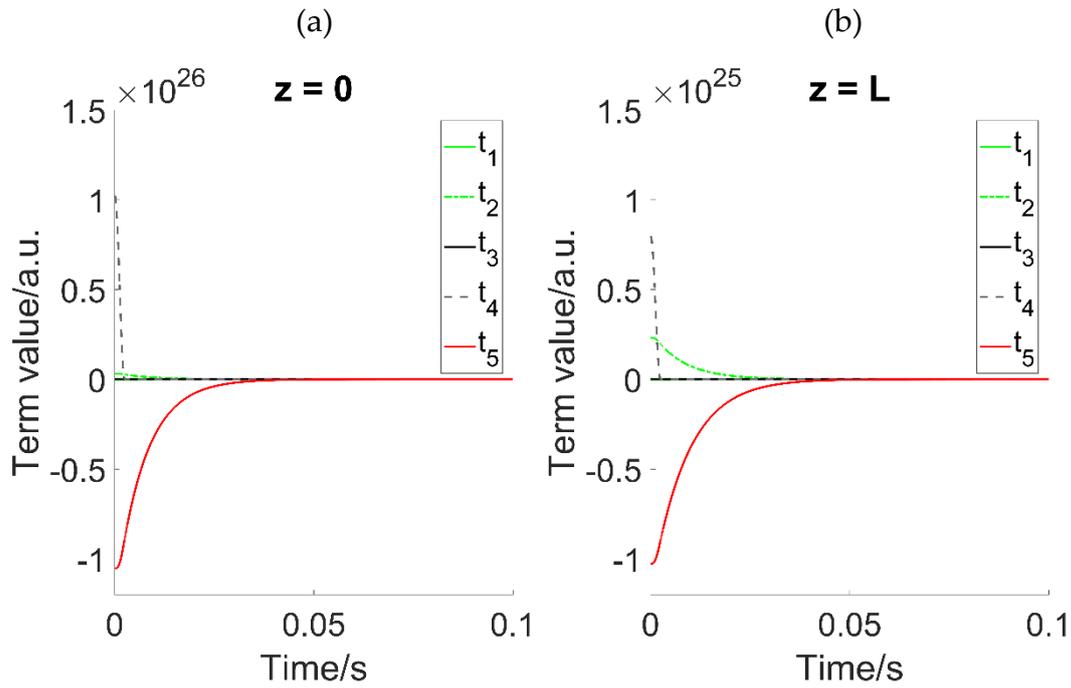

**Figure 7.** The time dependence of terms appearing on the right hand side of equation (1b) listed according to Table 1 at (a) z = 0 and (b) z = L.

These results from Fig.8 show that apart from a difference between the results calculated using the time dependence of $N_2$ at z = 0 and z = L, there is also a constant offset (difference from the reference value) for the time constant extracted from the time dependence of $N_2$. This offset does not decay to zero even at very large values of the observation time. Further, Fig.8b shows that after the initial few milliseconds during which the chopper switches off the pump the time constants extracted from the derivative of logarithm $N_2$ at z = 0 approach the reference value of $1/(1/\tau_{2r} + 1/\tau_{21nr})$ much closer than those calculated at z = L. Fig.8b shows that initially the time decay of $N_2$ recorded at z = 0 is better suited for extraction of the implicit material time constant that the one recorded at z = L. This observation is thus in agreement with the prediction made on the basis of the results shown in Fig.7. However, as seen in Fig.8a where the vertical axis is expanded to allow for a more accurate comparison, the good agreement between the time constant extracted from results recorded at z = 0 (Fig.8b) is to some extent deceptive due to coarse scaling of the vertical axis, and also there is a constant and non-vanishing offset between the lifetime extracted from the $N_2$ decay curve and the reference value for large values of the observation time. For large values of the observation time this offset is the same for both $N_2$ recorded at z = 0 and z = L. To further investigate this point the ratio of the absolute values of the terms $t_2$ and $t_5$ has been calculated and shown in Fig.9. The results from Fig.9 confirm that there is a constant ratio between both terms for large values of time, i.e. the contribution of $t_2$ does not vanish when compared with t₅ even for large values of the observation time. Thus there is no reason to expect that the dominance of term 5 when compared to term 2 will be increasing with time. Hence, a constant offset is to be expected and one can conclude that the extraction of photoluminescence decay lifetime $1/(1/\tau_{2r} +1/\tau_{21nr})$ from the $N_2$ decay curves is accompanied by an error resulting from a non-negligible contribution of stimulated absorption to the $N_2$ decay process.

Finally it is shown that a simple analytical formula can be derived for estimating the error caused by the contribution from the stimulated emission. For this purpose an approximation to equation (1b) is derived for large values of the observation time. From equation (1b) for a given z position one obtains by neglecting all terms except 2 and 5, and approximating $N_1$ by $N_{Tb}$, which is valid for large values of t, the following equation:

$$\frac{dN_2}{dt} \approx \overline{\sigma}_{21a} N_{Tb} \phi_s - \frac{N_2}{\tau_{2r}} - \frac{N_2}{\tau_{21nr}} \tag{5}$$

One can then derive from equations (3b) an approximate expression for the steady state value of $\phi_s$, valid for large values of time, which ignores stimulated emission term and spatial derivative term and yields a linear dependence between $\phi_s$ and $N_2$:

$$\phi_s = N_2 \eta / \tau_{2r} \left( \overline{\sigma}_{21a} N_{Tb} + \overline{\alpha}_s \right) \tag{6}$$

Combining (5) and (6) yields the equations describing the dynamics of $N_2$, which gives the following formula for the asymptotic value of the time constant:

$$\tau_{as} = 1/\left(1/\tau_{2r} + 1/\tau_{21nr} - \eta \mu / \tau_{2r}\right) \tag{7}$$

where $\mu = \overline{\sigma}_{21a} N_{Tb} / \left( \overline{\sigma}_{21a} N_{Tb} + \overline{\alpha}_s \right)$. For the case studied formula (7) yields a time constant of 7.63 ms, while the time obtained from Fig.8 after enlargement of the relevant area at time = 1 s is 7.59 ms, which shows a very good agreement. The reference value is 6.99 ms. So asymptotically this corresponds to an error of approximately 8 %. It is noted however, that (7) gives the upper bound for the error when collecting the results at z = 0 (cf. Fig.8a, excluding the initial few ms whilst the chopper switches off the pump). It is also important to note that in practice it is very difficult, if not impossible, to observe the photoluminescence decay for 1 s due to noise. In the experimental results presented in Fig.6 after 40 ms the curves practically 'sink' in noise. Hence only the first 30 ms can be used for extracting the time constant, which in practice would yield a lower error when using the photoluminescence recorded at z = 0. Formula (7) also indicates that the measurement error can be reduced by decreasing the doping concentration. This however, is accompanied by weakening the measured signal and increasing the noise. Also, according to formula (7), a fairly high loss helps improving the accuracy. However, this latter conclusion might be misleading since high loss at relevant wavelengths result usually from impurities [33], which are known in turn to modify the intrinsic lanthanide ion lifetimes [34]. Finally, reducing the spontaneous emission coupling coefficient η reduces the error according to (7). Thus making a structured core/clad fiber with a small η might help to reduce the measurement error . Otherwise, an application of a spatiotemporal model, as described in this contribution, allows for an extraction of the relevant material constants without an error imposed by the stimulated absorption effect.

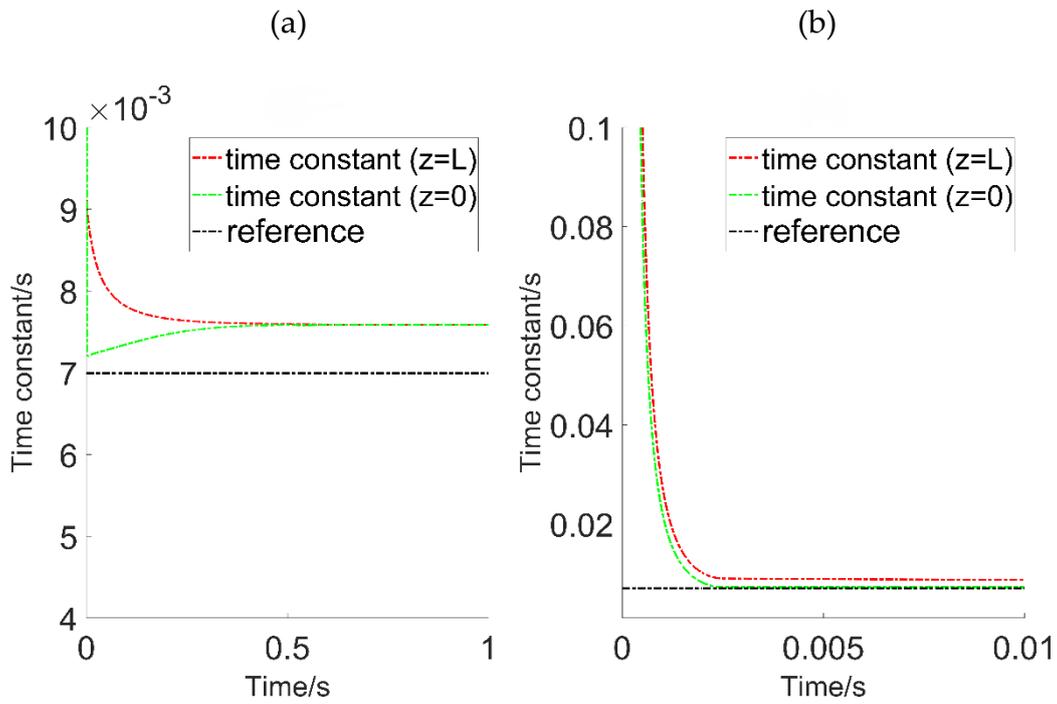

**Figure 8.** Photoluminescence decay time constant extracted numerically from the time dependence of $N_2$ (a) for time values up to 1 s and (b) for first 10 ms.

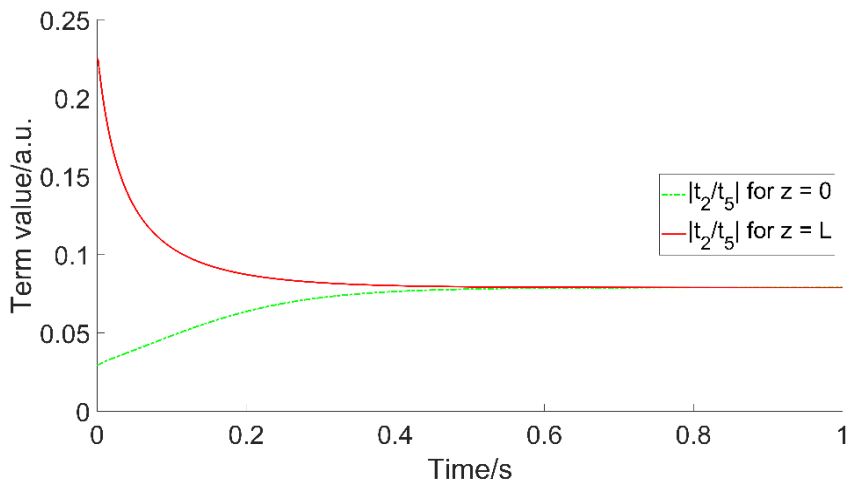

**Figure 9.** The time dependence of absolute values for the ration of $t_2$ and $t_5$ from Table 1 at z = 0 and z = L.

## 4. Conclusions

In summary, a detailed experimental and numerical analysis of photoluminescence lifetime measurement from lanthanide ion doped samples was performed. The discrepancies in the time dependence of photoluminescence collected from various positions along the fiber are due to the fact that apart from the spontaneous emission the stimulated absorption makes a significant contribution to the photon interactions between the relevant electronic levels. The results presented indicate that the stimulated absorption contribution cannot be eliminated even at large observation times. Thus a direct extraction of the material radiative constants from the photoluminescence decay curves is accompanied by an error, which is difficult to eliminate unless a spatiotemporal numerical model is applied numerically fit the observed photoluminescence decay.

**Acknowledgments:** This Project has received funding from the European Union's Horizon 2020 research and innovation programme under the Marie Skłodowska-Curie grant agreement No. 665778 (National Science Centre, Poland, Polonez Fellowship 2016/21/P/ST7/03666) 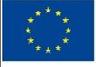. The authors wish to thank also Wroclaw University of Science and Technology (statutory activity) for financial support.